\documentstyle[twoside,fleqn,espcrc2]{article}
\def\be{\begin{equation}}
\def\ee{\end{equation}}
\def\bea{\begin{eqnarray}}
\def\eea{\end{eqnarray}}

\title{ Charmonium Spectroscopy From
Lattice NRQCD }

\author{NRQCD Collaboration presented by A.J. Lidsey
               \address{Department of Physics and Astronomy,
           University of Glasgow, \\
           Glasgow, G12 8QQ, Scotland, UK}}

\begin{document}

\begin{abstract}
We present the first set of results for Charmonium Spectroscopy
using Non-Relativistic QCD (NRQCD). For the NRQCD action
the leading order spin-dependent
and next to leading order spin-independent interactions
have been included. Clear signals for the s and p hyperfine splittings
have been observed as well as various orbital states.

\end{abstract}

\maketitle

\section{Introduction}
NRQCD is an effective field theory where a cut off of the order of
the quark mass is introduced into QCD in order to remove the mass scale
from the calculation.
Both numerical and theoretical
work has been done in developing NRQCD on the lattice
where the lattice
spacing provides the cut off.
The NRQCD action is essentially an expansion of the usual QCD action in powers
of the individual quark velocities squared $\frac{v^{2}}{c^{2}}$ , where for
the bottom
and charm quarks $\frac{v^{2}}{c^{2}} \approx $ 0.1 and 0.3 respectively.
For our action we use the lagrangian defined in Euclidean
space by
\bea \label{LNRQCD}
{\cal{L}}_{NRQCD} = -\Psi^{\dagger}D_{t}\Psi + \Psi^{\dagger}\frac{D^{2}}{2M
_{Q}}\Psi + \nonumber \\
\delta{\cal{L}}_{SI} + \delta{\cal{L}}_{SD}
\eea
where
\bea \label{LSI}
{\cal{L}}_{SI} =
-c_{2}i\frac{g}{8M_{Q}^{2}}\Psi^{\dagger}(D.E - E.D)\Psi + \nonumber \\
c_{1}\frac{1}{8M_{Q}^{3}}\Psi^{\dagger}D^{4}\Psi
\eea
are the next to leading order spin-independent terms, and
\bea \label{LSD}
{\cal{L}}_{SD} =
c_{3}\frac{g}{8M_{Q}^{2}}\Psi^{\dagger}
\sigma.(D^{\wedge}E - E^{\wedge}D)  + \nonumber \\
c_{4}\frac{g}{2M_{Q}^{2}}\Psi^{\dagger}\sigma.B\Psi
\eea
are the leading order spin-dependent terms. Since we have included next to
leading order
spin-independent terms it is possible when using
charm quarks to obtain an accuracy of $\approx10\%$ for
quantites such as the spin averaged S-P splitting
or the 2S-1S splitting. For spin-dependent quantities
like $^{3}S_{1}-^{1}S_{0}$ an accuracy of only
$\approx 30\%$ is possible.
\\
To define ${\cal{L}}_{NRQCD}$ on the lattice \cite{NRQCDcolab}
covariant derivatives are represented
by shift operators
\be \label{shift}
\triangle^{+}_{i}\Psi(x) = U_{\mu}(x)\Psi(x+i) - \Psi(x)
\ee
and the laplacian is defined by
\be \label{laplacian}
\triangle^{2} = \sum_{i} \triangle^{+}_{i}\triangle^{-}_{i}
\ee
To evaluate the chromomagnetic and electric fields the usual cloverleaf
expression is used to find F$_{\mu\nu}(x)$ and hence ${\bf E(x)}$ and
${\bf B(x)}$.
The arbitary constants appearing in eq. (\ref{LSI}) and (\ref{LSD}) are
evaluated by
matching NRQCD to full QCD at low energies. At tree level
all the c's are unity
\cite{cornell}. The biggest correction coming from loop
diagrams will be tadpole corrections arising due to the non-linear
relation between the gluon field and the lattice gluon operator. In order
to remove this we use the method suggested by \cite{lepmac} where all the U's
are redefined by
\be \label{US}
U_{\mu}(x) \rightarrow \frac{U_\mu(x)}{u_{0}}
\ee
with $u_{0} = <0|\frac{1}{3}TrU_{PLAQ}|0>^{\frac{1}{4}}$. Since the
cloverleaf expression involves the evaluation
of a plaquette this
renormalization will have the effect of redefining
${\bf E}$ and ${\bf B}$ via
\be \label{EB}
{\bf E } \rightarrow \frac{{\bf E}}{u^{4}_{0}}  \hspace{15pt}
{\bf B } \rightarrow \frac{{\bf B}}{u^{4}_{0}}
\ee
which will strongly affect spin-dependent quantities. Not removing
these tadpole contributions will result in an underestimation of
the electric and magnectic fields and hence hyperfine splittings.
The tadpoles are the dominant affect in radiative corrections
so removing them enables us to use the tree level values
for the c's.
\\
As with all lattice simulations lattice spacing errors will need to be
investigated
and corrected for. In NRQCD for heavy-heavy quark systems correction terms can
be added as a power series in the typical quark velocity in the same way
as relativistic corrections were in eq. (\ref{LSI}) and eq. (\ref{LSD}).
In order to obtain the desired
accuracy quoted above it is necessary to remove $O(a^{2})$ corrections
from the laplacian by redefining $\triangle^{2}$ as
\be \label{spcorec}
\tilde{\triangle}^{2} = \sum_{i} \triangle^{+}_{i}\triangle^{-}_{i} -
\frac{a^{2}}{12}\sum_{i}[\triangle^{+}_{i}\triangle^{-}_{i}]^{2}
\ee
and for the temporal correction it is sufficient to redefine the kinetic
operator
as
\be \label{tcorec}
-\frac{\tilde{\triangle^{2}}}{2M} \rightarrow -\frac{\tilde{\triangle^{2}}}{2M}
-\frac{a}{4n}\frac{(\triangle^{2})^{2}}{4M^{2}}
\ee
which still allows the evolution equation to be an initial value problem.
n is an even number introduced to prevent instabilities occurring in the
simulation at low
quark mass ~\cite{thaclep,cornell}.
\\
\section{\bf Simulation results}
In the simulation we use a lattice of size $12^{3}$x24 at $\beta = 5.7$
with quenched gauge configurations supplied by the UKQCD collaboration, fixed
to Coulomb gauge. The operators are smeared using a coulombic wavefunction
as a smearing function.
Due to  the relatively small size of charmonium it is possible to use
more than one starting site on the lattice and since our evolution equation
is a simple difference one, different starting times can be used too.
For our particular lattice we use 8 different spatial origins and 2 different
starting times.
\\
For NRQCD the dispersion relation has the form
\be \label{dispers}
E_{p} = M_{1} + \frac{p^{2}}{2M_{2}} + ....
\ee
where, because we have removed the $M_{Q}\Psi^{\dagger}\Psi$ term from the
original Lagrangian, $M_{1}\neq M_{2}$.
$M_{1}$ is a redundent zero of energy and it is $M_{2}$ which
is the important dynamical mass determining spin averaged
splittings. The bare quark mass $M_{Q}$
appearing in the Lagrangian is tuned in the simulation
so that $M_{2}$ appearing in the dispersion relation
for the $^{1}S_{0}$ state
is equal to the experimental value of the $\eta_{c}$. This
is done by Fourier transforming the correlation function
into momentum space from which $E_{p_{min}}$
can be found, where $p_{min}$ is
the lowest allowed non-zero
momentum on the lattice. Using eq. (\ref{dispers}) $M_{2}$ is
evaluated from $E_{p_{min}}-E_{0}$.
\\
In our simulation we have used 100 configurations and
a bare quark mass in lattice units of 0.8. Extracting the lattice spacing from
the spin
averaged S-P splitting, a value known to be insensitive
to the bare quark mass $M_{Q}$,
a value of $a^{-1}$ = 1.23 (7) GeV is obtained. Using
this we find $M_{2}$ = 3.00 Gev for the $^{1}S_{0}$ to be
compared to the experimental value
of 2.98 GeV.
\\
Shown in figure (\ref{figSI})
are values for various spin-independent
splittings
where mostly single exponential fits have been used.
\begin{table}
\begin{tabular}{|r|c|l|}
\hline
Splitting & Lattice result  & Exp \\
          &   (MeV)         & (MeV) \\
\hline
$^{3}S_{1}-^{1}S_{0}$ & 96 (4) & 117 \\ \hline
$^{3}P_{2}-^{3}P_{1}$ & 60 (30) & 46 \\ \hline
$^{3}P_{1}-^{3}P_{0}$ & 52 (9) & 95 \\ \hline
\end{tabular}
\vspace{9pt}
\caption{Results for s and p hyperfine splittings
for the 1s and 1p states.}
\label{tabSD}
\end{table}
\begin{figure}[htb]
\vspace{75mm}
\caption{Results for spin-averaged splittings
plotted relative to the spin-averaged 1S state.
The vertical scale is in GeV. The horizontal lines mark experimental
results, the squares are results from this simulation. Crosses represent
splittings which were fixed to experiment.
The $\psi(3770)$ is a $^{3}D_{1}$ state compared to our result for the
$^{1}D_{2}$ (not spin-averaged).}
\label{figSI}
\end{figure}
The $^{1}D_{2}-^{1}S_{0}$ splitting has to be compared to the
experimental $^{3}D_{1}-^{1}S_{0}$ result  since the $^{1}D_{2}$ has
not been observed experimentally.
The $^{3}D_{1}$ has the same quantum numbers as the $^{3}S_{1}$
and will appear as a third excited state state in that channel making it
difficult to extract a value for its mass.
The simulation result is slightly
higher as one would expect and certainly in the right area considering the
$^{1}
D_{2}$ is near the threshold for decay into D mesons.
For the
2S-1S splitting the 2S was obtain by performing a three
exponential fit to two correlation functions. The result
matches the experiment although the error bar is large.
\\
Tabel (\ref{tabSD}) shows values for spin-dependent splittings.
For the S hyperfine splitting the value is well
within the expected systematic accuracy even though it is smaller
than the experimental value.
Since this splitting is caused by a local-local interaction the reduction
might be caused by quenching. This affect has been estimated to cause a
reduction of $\approx 20 \%$ \cite{cthd}.
The fact that the s hyperfine is within the
expected accuracy strongly suggests that the improved tadpole
coupling constant in the $\sigma.B$ interaction
is the correct value to use.
For the p hyperfines the statistical
errors have increased but still a signal can be seen. The
agreement with experiment
suggests that the $\sigma .D^{\wedge}E$ term can correctly account
for the p hyperfine splitting, again using a tadpole improved coupling.
\section{Conclusion}
Using NRQCD it is possible to obtain a spectrum for
Charmonium in good
agreement with experiment.

In the future more configurations will be used to reduce statistical errors.
This will provide a stringent test of our perturbatively improved action
and so of QCD itself.
Using perturbation theory with the numerical results will then
enable a value
for the charm quark mass to be determined.
\section{Acknowledgements}
A.J. Lidsey and C.T.H Davies
would like to thank the UK SERC for supporting this work
and the UKQCD collaboration
for supplying the configurations.
They were generated on a Meiko i860 computing surface
supported by SERC grant GR/G32779, Meiko Limited, and
the University of
Edinburgh. The computer simulations presented here were
performed on the ymp8
at the Ohio Supercomputer centre and at the Atlas centre RAL, UK.

\end{document}